\documentclass[pra,showpacs,preprintnumbers,amsfonts,amssymb,amsmath,floats,twocolumn,aps]{revtex4}

\usepackage{graphicx}
\usepackage{dcolumn}
\usepackage{bm}
\usepackage{color}

\newcommand{\ff}[1]{{\boldsymbol #1}}

\newcommand{\ba}{\begin{eqnarray}}
\newcommand{\ea}{\end{eqnarray}}
\newcommand{\be}{\begin{equation}}
\newcommand{\ee}{\end{equation}}

\begin{document} 
  
\title{Cooperation of different exchange mechanisms in confined magnetic systems}
\author{Andrej Schwabe, Mirek H\"ansel and Michael Potthoff}
\affiliation{I. Institut f\"ur Theoretische Physik, Universit\"at Hamburg, Jungiusstra\ss{}e 9, 20355 Hamburg, Germany}

\begin{abstract}
The diluted Kondo lattice model is investigated at strong antiferromagnetic local exchange couplings $J$, where almost local Kondo clouds drastically restrict the motion of conduction electrons, giving rise to the possibility of quantum localization of conduction electrons for certain geometries of impurity spins. This localization may lead to the formation of local magnetic moments in the conduction-electron system, and the inverse indirect magnetic exchange (IIME), provided by virtual excitations of the Kondo singlets, couples those local moments to the remaining electrons. Exemplarily, we study the one-dimensional two-impurity Kondo model with impurity spins near the chain ends, which supports the formation of conduction-electron magnetic moments at the edges of the chain for sufficiently strong $J$. Employing degenerate perturbation theory as well as analyzing spin gaps numerically by means of the density-matrix renormalization group, it is shown that the low-energy physics of the model can be well captured within an effective antiferromagnetic RKKY-like two-spin model (``RKKY from IIME'') or within an effective central-spin model, depending on edge-spin distance and system size.
\end{abstract} 
 


\maketitle 

\section{Introduction}

It has recently been suggested \cite{GHG+10,ZBL+14,SHH+14} that fermionic alkaline-earth atoms 
\cite{DTH+13} can be used to efficiently simulate condensed-matter systems with spin and 
orbital degrees of freedom.
One of those many-body systems is the one-dimensional Kondo-lattice model \cite{TSU97b} with 
its intricate interplay between Kondo screening and magnetic order \cite{FFHR10}.
Particularly, the regime of strong antiferromagnetic local exchange coupling $J$ is accessible to the
experiment \cite{GHG+10,ZBL+14,SHH+14}.
The purpose of the present paper is to demonstrate that quantum confinement effects resulting from 
strong $J$ in a {\em spin-diluted} system can effectively result in a {\em weak} indirect magnetic
RKKY \cite{RK54,Kas56,Yos57} interaction.
This is achieved by exploiting the characteristics of a novel ``inverse'' indirect magnetic 
exchange (IIME) mechanism that has been proposed recently \cite{STP13}.
We study the distance dependence of the effective IIME coupling in a one-dimensional prototypical two-impurity model by means of 
strong-coupling perturbation theory and density-matrix renormalization group (DMRG).
We argue that the spin-diluted Kondo lattice opens a new field where the complex, 
cooperative as well as competitive interplay between the Kondo effect \cite{Hew93} and different 
kinds of indirect magnetic exchange mechanisms can be studied in quantum-confined 
geometries.

The conventional RKKY exchange between two impurity spins is mediated by itinerant conduction electrons and leads to an effective, indirect magnetic coupling $J_\text{RKKY}$. This coupling is obtained from second-order perturbation theory in the local exchange interaction $J$ between the local spins and the conduction electrons. It is oscillatory in the distance $d$ of the spins, $J_\text{RKKY} \sim (-1)^d J^2 / d$, for a non-interacting one-dimensional metallic host system given by a tight-binding model with nearest-neighbor hopping $t$ at half-filling. On the other hand, if $J$ is much larger than $t$, a strong-coupling variant of RKKY exchange \cite{TDHFP13,SA96} can be derived perturbatively in powers of $t/J$. If $J$ is antiferromagnetic, RKKY exchange typically competes with the emergence of the Kondo effect \cite{Don77} which is responsible for the individual screening of impurity spins by the conduction electrons. The Kondo temperature $T_K$ is the corresponding energy scale of the crossover to the screened regime and can be converted by the Fermi velocity $v_F$ into a length scale $\xi_K\sim v_F/T_K$ which may be interpreted as the extension of a Kondo screening cloud. The weak-coupling regime is dominated by RKKY exchange, while the Kondo regime is realized at strong couplings $J$.

Here, we consider the competition between Kondo effect and RKKY exchange in \emph{confined} systems with open boundaries.  I.e., rather than discussing effects of the particular form of a trapping potential, infinitely large potential barriers are assumed for simplicity. 
The spectral gap $\Delta$ of the confined, non-interacting host system near the Fermi edge acts as a cutoff for the characteristic Kondo correlations\cite{TKvD99,SA03,TSRP12} in case of $\Delta>T_K$. 
This also considerably affects the competition between the RKKY exchange and the Kondo effect \cite{SGP12} and can lead to exotic ground states and even to a reappearance of a Kondo regime for $J\rightarrow0$.

Unconventional finite-size effects are also found in the strong-coupling limit.
Although the Kondo effect quenches the impurity spins for $J\gg t$, it has been realized that it may also \emph{help} to generate magnetism in the case of systems with diluted impurity spins \cite{STP13}, or, more generally, diluted correlated impurity sites \cite{TSP14}. Namely, almost local Kondo singlets can result in an efficient additional confinement of the itinerant electrons. This implies the formation of local moments in the otherwise non-interacting conduction electron system. Moreover, these local moments are coupled magnetically by virtual excitations of the magnetically inert Kondo singlets. 
This constitutes an ``inverse'' indirect magnetic exchange (IIME) \cite{STP13} where the roles of the conduction electrons and of the impurity spins are essentially interchanged.

The strong-coupling Kondo physics is necessarily local or almost local and induces, via the IIME mechanism, a short-range coupling between local conduction-electron magnetic moments.
As is shown here, the IIME mechanism also applies to situations where the additional confinement extends over larger regions.
An interesting question is if this can be employed to generate a situation where effective RKKY-like interactions couple magnetic moments over larger distances:
Can one make use of the strong-coupling limit and of a strong and almost local Kondo effect to implement an effective weak-coupling model free of any competing Kondo screening?
How does the effective magnetic coupling depend on the distance between the moments and on the bare coupling $J$?
These questions are actually part of a more general approach to understand the complex interplay between magnetic exchange interactions and the Kondo effect in quantum-confined systems.

\section{Two-impurity model in the strong-coupling limit}

\begin{figure}[b]
\centerline{\includegraphics[width=0.3\textwidth]{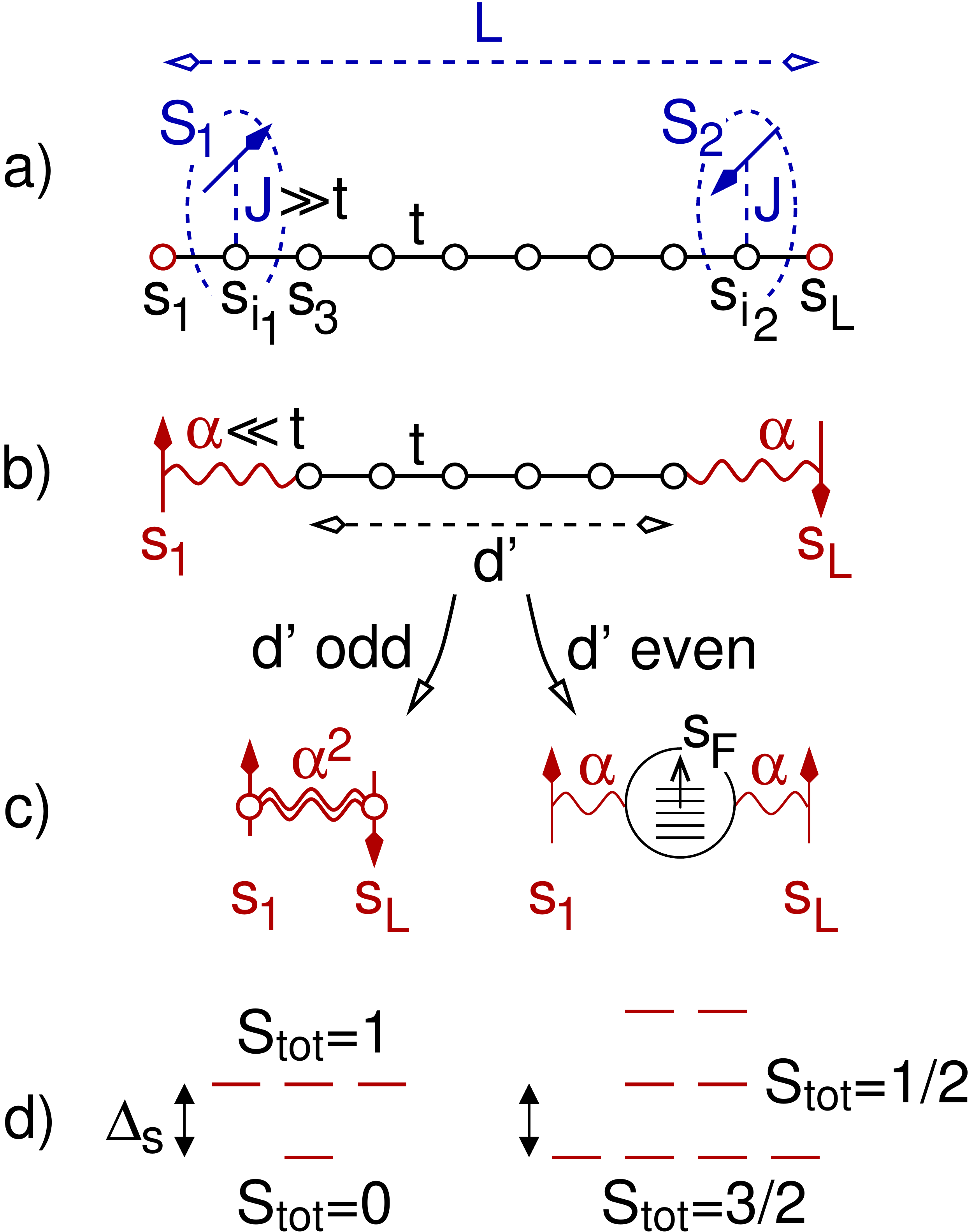}}
\caption{(Color online)
a) Two-spin Kondo model: Impurity spins, $\ff S_{1}$ and $\ff S_{2}$ at a distance $d=L-3$ couple via a strong antiferromagnetic exchange $J$ to the local spins $\ff s_{i_{1}}$ and $\ff s_{i_{2}}$ at sites $i_{1}=2$ and $i_{2}=L-1$ of a one-dimensional chain of $L$ sites. $t$ is the nearest-neighbor hopping in the half-filled system of conduction electrons.
b) Effective model for $J\gg t$: The impurity spins form local Kondo singlets with $\ff s_{i_{1}}$ and $\ff s_{i_{2}}$. This leads to local-moment formation at the edge sites $i=1$ and $i=L$. An ``inverse'' indirect magnetic exchange with coupling constant $\alpha$ emerges and mediates a ferromagnetic coupling of the edge spins $\mathbf s_1$ and $\mathbf s_L$ to the rest of the conduction-electron sea. 
c) Effective RKKY and effective central-spin model \cite{G76}: For even $L$, i.e. for odd distance $d^\prime=L-5$ between the edge spins in the effective model, the low-energy model is a two-spin antiferromagnetic RKKY model with RKKY exchange $\alpha^2$.
For odd $L$ ($d^\prime$ even), the low-energy model is a central-spin model where $\ff s_{1}$ and $\ff s_{L}$ couple ferromagnetically with strength $\alpha$ to the (delocalized) spin $\ff s_{\rm F}$ of the singly occupied Fermi orbital $k_{\rm F}$.
d) Low-energy spectrum: For even $L$ (odd $d^\prime$), the ground state is a spin singlet, $\Delta_{\rm s}$ is the singlet-triplet excitation energy. For odd $L$ (even $d^\prime$), the ground state is a spin quartet. For non-interacting conduction electrons this quartet is degenerate with a doublet.
}
\label{fig:model}
\end{figure}

To study those questions, we consider a one-dimensional prototype model as depicted in Fig.\ \ref{fig:model}a. Two impurity spins $\ff S_{1}$ and $\ff S_{2}$ (with quantum number $S=1/2$) are strongly coupled locally via an antiferromagnetic exchange $J\gg t$ to the local conduction-electron spins $\ff s_{i_1}=\ff s_{2}$ and $\ff s_{i_2}=\ff s_{L-1}$ of a system of $N$ itinerant and non-interacting conduction electrons. The conduction-electron system is half-filled, i.e. $N=L$ where $L$ is the number of lattice sites. The two resulting local Kondo singlets at strong $J$ are located at a distance $d=L-3$, thereby defining an intermediate ``central region'' of sites $i=3,\ldots,L-2$. The hopping of the conduction electrons is $t_{ij} = -t$ between non-degenerate orbitals on nearest-neighboring sites $i,j$ of the lattice. Throughout the paper, all energies are given in units of the nearest-neighbor hopping, i.e., $t=1$ fixes the energy scale.

The Hamiltonian is 
\begin{equation}
{H} = \sum_{ i,j, \sigma} t_{ij} c^{\dagger}_{i\sigma} c_{j\sigma} 
+ 
J \sum_{m=1}^{2} \ff s_{i_{m}} \ff S_{m} 
\: .
\label{eq:ham}
\end{equation}
Here, $c_{i\sigma}$ annihilates an electron at site $i=1,...,L$ with spin projection $\sigma=\uparrow, \downarrow$, and 
$\ff s_{i} = \frac{1}{2} \sum_{\sigma \sigma'} c^{\dagger}_{i\sigma} \ff \sigma_{\sigma\sigma'} c_{i\sigma'}$ is the local conduction-electron spin at $i$, where $\ff \sigma = \sum_{\alpha} \sigma^{\alpha} \ff e_{\alpha}$ is the vector of Pauli matrices and $\alpha=x,y,z$.

Let us first concentrate on even lattice sizes $L=4,6,\ldots$. Systems with odd $L$ are discussed in Sec. \ref{sec:ferro_dist}.
We also focus on the weak-coupling limit first.
For $J\rightarrow0$, the ground state is a total spin singlet ($S_\text{tot}=0$). Here, the Kondo correlations are cut by the finite system size\cite{TKvD99,SA03,TSRP12,SGP12}, i.e., $\Delta>T_K$, since the Kondo temperature is exponentially small, $T_K\sim e^{-1/J}$. The chemical potential $\mu$ falls into the finite-size gap $\Delta$ between fully occupied and unoccupied energy levels of the non-interacting conduction-electron system. Consequently, the impurity spins $\mathbf S_1$ and $\mathbf S_2$ are effectively decoupled from the host system at low energies. Using perturbation theory in $J$, one then obtains an RKKY Hamiltonian
\begin{align}
   \label{eq:rkky_ham}
   H_\text{RKKY} = J_\text{RKKY} \mathbf S_1\mathbf S_2
\end{align}
featuring an antiferromagnetic coupling $J_\text{RKKY}=J^2\chi_{i_1i_2}^{0,\text{cond}}(\omega=0)>0$, where $\chi_{i_1i_2}^{0,\text{cond}}(\omega=0)$ is the nonlocal static susceptibility of the non-interacting conduction-electron system. One finds that $J_\text{RKKY}\sim J^{2}(-1)^{d+1} /d$ for a one-dimensional system, where $d=L-3$ is the distance between the impurities.

However, in the strong-coupling regime $J\rightarrow\infty$, Eq. \eqref{eq:rkky_ham} cannot be the effective low-energy Hamiltonian anymore, since the impurity spins are compensated by individual Kondo effects and $J_\text{RKKY}\rightarrow0$. Perturbation theory, however, allows to study the remaining inter-impurity ground-state correlations for $J\gg t$, yielding the envelope function \cite{TDHFP13}
\begin{align}
    \label{eq:str_coup_rkky}
    \langle \mathbf S_1\mathbf S_2\rangle \sim 1/d^2
    .
\end{align}
This result can be regarded as a strong-coupling variant of RKKY theory. Furthermore, it can be interpreted as a signature of the fact that each impurity is located in the exterior of the Kondo cloud of the other impurity, as the low-energy physics of the single-impurity Kondo model is essentially the one of a Fermi liquid with \cite{I78,SA96,BA98} $\langle \mathbf S_m \mathbf s_j\rangle \sim 1/|i_m-j|^2$ for $|i_m-j|\gg\xi_K$.

One may explicitly derive the effective low-energy Hamiltonian by using strong-coupling degenerate perturbation theory, where hopping  to and off a local Kondo singlet is treated as a weak perturbation. The resulting effective IIME Hamiltonian excludes the (high-energy) local Kondo singlets at $i_2$ and $i_{L-1}$ and operates on the remaining conduction-electron degrees of freedom only \cite{STP13,TSP14}. Up to fourth order in $t/J$, for impurity distances $d\geq2$, and disregarding unimportant constant energy shifts, one obtains
\begin{eqnarray}
 \label{eq:eff_ham_iime}
 H_\text{eff}  
 &\sim& 
             H_t
 \nonumber \\
 & + &
              \frac{\alpha}{2}
              \sum_{i=1,3,L-2,L}
              \left(n_{i\uparrow} -\frac12\right) \left(n_{i\downarrow } -\frac12\right)
 \nonumber \\
 & + &   
              \alpha
              \sum_{(i,j)=(1,3),(L-2,L)}
              \Big[
              \mathbf t_i
	      \mathbf t_j
              -
              \mathbf s_i
	      \mathbf s_j
 \nonumber \\
 & - &
              \frac12
              \sum_\sigma
			   (
               c^\dagger_{i\sigma} 
			   c_{j\sigma} 
			   +
               c^\dagger_{j\sigma} 
			   c_{i\sigma} 
			   )
                            (
			   1
			   -
			   n_{i -\sigma} 
			   -
			   n_{j -\sigma} 
			   )
              \Big]
              ,
 \nonumber \\
\end{eqnarray}
where
\begin{align}
                \label{eq:iime_alpha}
		\alpha = \frac{64}{3} \frac{t^4}{J^3}
\end{align}
is the IIME coupling constant and
\begin{align}
     H_t = \sum_{i,j=3}^{L-2}  \sum_\sigma  t_{ij} c^{\dagger}_{i\sigma} c_{j\sigma} 
\end{align}
the tight-binding Hamiltonian of the remaining conduction-electron sea of the central region. Furthermore, $n_{i-\sigma}=c^\dagger_{i-\sigma} c_{i-\sigma}$ is the particle number of electrons with spin projection opposite to $\sigma$, and $\mathbf t_i$ denotes the isospin \cite{UTS92} at site $i$, defined as $\mathbf t_i =\frac12\left( c^\dagger_{i\uparrow},(-1)^ic_{i\downarrow} \right) \cdot \pmb \sigma \cdot \left( c_{i\uparrow},(-1)^ic^\dagger_{i\downarrow} \right)^T$. Together with the SU(2) spin symmetry, the isospin SU(2) symmetry constitutes the SO(4) symmetry of the effective Hamiltonian at half-filling.

\section{Formation of edge spins}

Analyzing $H_\text{eff}$ (Eq. \eqref{eq:eff_ham_iime}), one may imagine two different scenarios for the low-energy physics. On the one hand, we may expect the formation of non-magnetic edge isospins or charge fluctuations at the edge sites. On the other hand, it also appears plausible that stable magnetic edge spins evolve for increasing $J\gg t$.

The interplay of the antiferromagnetic isospin coupling and the ferromagnetic spin coupling (third and fourth term in Eq. \eqref{eq:eff_ham_iime}) alone would result in the formation of isospins at the edge sites $i=1$ and $i=L$. This is prevented, however, by the repulsive local Hubbard-like interaction term (second term) which favors the formation of magnetic moments at the edges and suppresses doubly occupied or empty sites, i.e.\ the formation of isospins. Nonetheless, the formation of magnetic edge spins could still be hampered by the charge fluctuations triggered by the spin-isospin interaction term (fifth term).

We have performed exact diagonalization calculations for small lattices $L\leq10$. It is found that the first scenario is not realized, since $\langle \ff s_1^{2} \rangle \to 3/4$ for $J\rightarrow\infty$, e.g. $\langle \ff s_1^{2} \rangle \approx 0.74$ for $J=10$ and $L=8$. This shows that the isospin exchange term and the spin-isospin interaction term in Eq. \eqref{eq:eff_ham_iime} are not relevant for the low-energy spectrum of our model. Hence, we may think of it as an effective model with two $S=1/2$ edge spins (see Fig. \ref{fig:model}b) which are ferromagnetically coupled with a weak effective interaction strength $\alpha$ to the remaining electrons of the central region. An RKKY-like mechanism may then be responsible for an indirect coupling of both spins $\mathbf s_1$ and $\mathbf s_L$, resulting in an effective antiferromagnetic two-spin model due to their odd distance $d^\prime=L-5$. In fact, the exact-diagonalization calculations for $L<10$ predict a total spin-singlet ground state.

\section{Effective RKKY model for even $L$}

To gain further analytical understanding of the edge spin coupling, one would ideally have to perform an eighth-order strong-coupling perturbation theory. This quite demanding task can be circumvented by taking the effective Hamiltonian $H_\text{eff}$ as starting point for a second renormalization step and performing standard perturbation theory \cite{EFG+05} in powers of $\alpha/t$ since $\alpha\ll t$. 

One should note that the onsite interaction in $H_\text{eff}$ (second term) makes the central region correlated at the interfaces between the Kondo singlets and the central region, i.e. $i=3$ and $i=L-2$. Since the spins, developing at the chain edges for strong $J$, are only weakly coupled to the central region ($\alpha\ll t$), finite-size effects play an important role at sufficiently large $J$: The finite-size gap $\Delta^\prime \propto t$ of the remaining conduction-electron system near the Fermi energy is the largest energy scale  appearing in $H_\text{eff}$ (Eq. \eqref{eq:eff_ham_iime}). Consequently, we may safely neglect the additional interactions at the interface sites for a moment, as they become relevant only on much smaller energy scales $\sim \alpha$ and will give small corrections only.

With this idea, the unperturbed Hamiltonian is given by $H^\prime_0=H_t=\sum_{\langle i,j \rangle=3}^{L-2} \sum_\sigma  t_{ij} c^{\dagger}_{i\sigma} c_{j\sigma} $ and the perturbation by $H^\prime_1=-\alpha(\mathbf s_1 \mathbf s_3+\mathbf s_{L-2}  \mathbf s_L)$. Single-particle excitations of the non-magnetic central region require at least an energy of $\Delta^\prime$, i.e., this is a so-called off-resonance situation (see Ref.\ \cite{SGP12}). Hence, there is no contribution linear in $\alpha$ and second-order perturbation theory predicts an effective RKKY model \cite{TSRP12,SGP12}
\begin{align}
    \label{eq:eff_ham_iime_offr}
    H_\text{eff}^\prime = J^\prime \mathbf s_1 \mathbf s_L
    ,
\end{align}
illustrated in the left panel of Fig. \ref{fig:model}c.

The RKKY exchange coupling is given by
\begin{align}
    \label{eq:delta_s_offr_pt}
    J^\prime = -
               \frac{\alpha^2}{2} 
               \sum_{\varepsilon_k<\mu,\varepsilon_p>\mu} 
               \frac{1}{\varepsilon_p - \varepsilon_k} 
               U_{3,k}U_{3,p}U_{L-2, p}U_{L-2, k}
    ,
\end{align}
where $U_{i,k}$ is the local weight of the state with momentum $k$ of the remaining conduction-electron sea at site $i$ and $\varepsilon_k$ its energy.

$J^\prime$ has the typical Fermi-liquid dependence \cite{LD98} on $\alpha$ and $d^\prime$ at large distances
\begin{align}
  \label{eq:iime_coupling}
  J^\prime\sim (-1)^{d^\prime+1} \alpha^2 / d^\prime 
          \sim \frac{ (-1)^{d^\prime+1} } { d^\prime } \frac{t^8}{J^6}
\end{align}
and is antiferromagnetic as the edge-spin distance $d^\prime=L-5$ is assumed as odd in our model \eqref{eq:ham}. The corresponding energy spectrum of $H_\text{eff}^\prime$ is shown in the left panel of Fig. \ref{fig:model}d, containing a singlet ground state and excited triplet states.

\section{DMRG calculations}

\begin{figure}[t]
\centerline{\includegraphics[width=0.95\columnwidth]{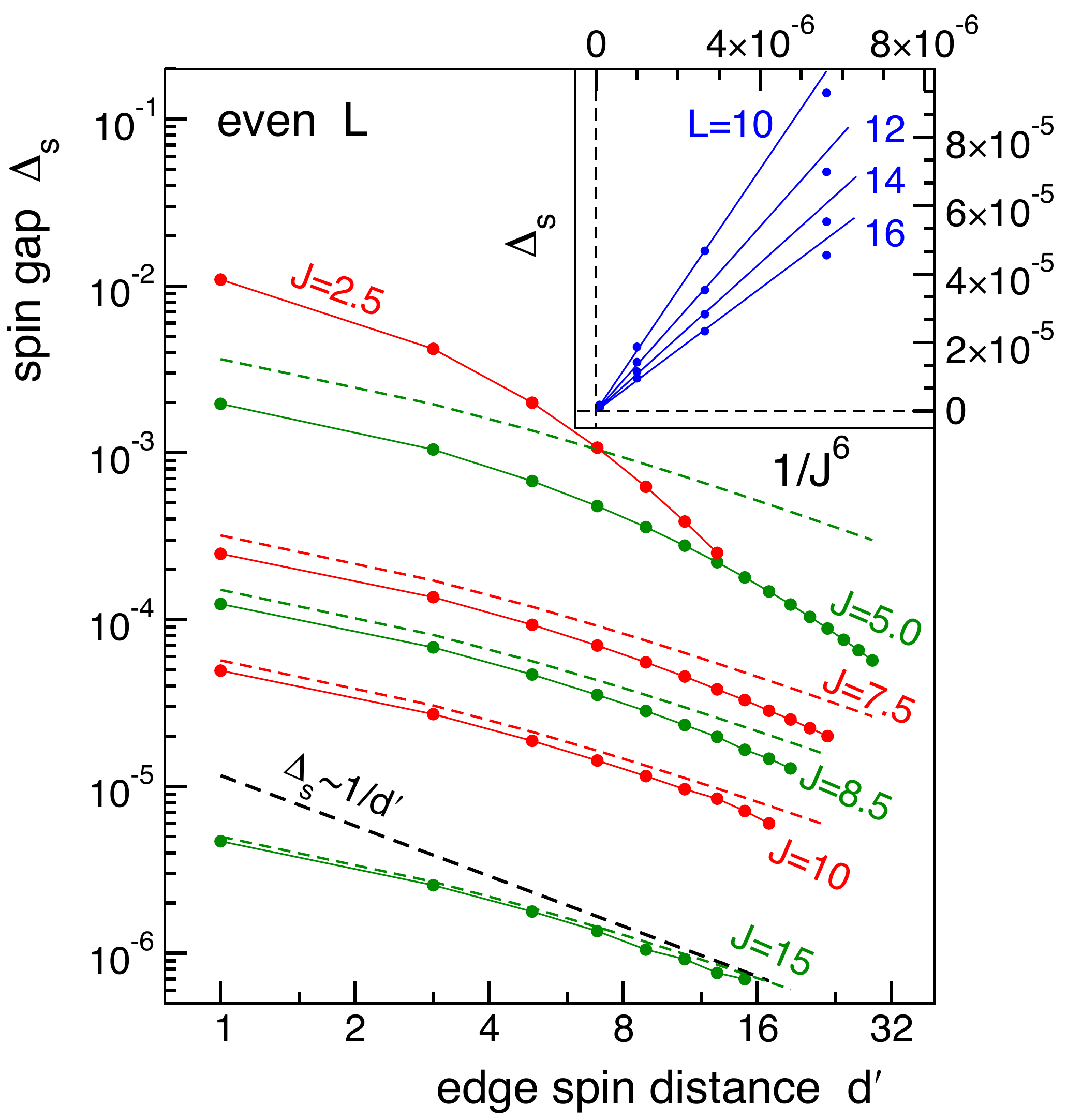}  }
\caption{(Color online)
Spin gap $\Delta_{\rm s}$ as a function of the edge spin distance $d^\prime$ on a double-logarithmic scale for different couplings $J$ as indicated. 
Dots: DMRG results for even $L$ and odd edge spin distance $d^\prime=L-5$, i.e.\ antiferromagnetic coupling. 
Colored dashed lines: Spin gap as given by second-order perturbation theory in $\alpha/t$, see Eq. \eqref{eq:delta_s_offr_pt}. Black dashed line: Expected $d^\prime$-dependence of $\Delta_s$ for large $d^\prime$ and large $J$.
Inset: $J$-dependence of the spin gap. For different system sizes $L$, the DMRG data follow the expected power law $\Delta_{\rm s} \propto 1/J^{6} \propto \alpha^{2}$ at strong $J$.
The nearest-neighbor hopping $t=1$ fixes the energy scale.
}
\label{fig:gap}
\end{figure}

For a numerical check of the distance and $J$ dependence predicted by the proposed effective Hamiltonian $H_\text{eff}^\prime$ (Eqs. \eqref{eq:eff_ham_iime_offr} and \eqref{eq:iime_coupling}), we apply a standard implementation of density-matrix renormalization group\cite{Whi92,Sch11} (DMRG) based on matrix product states and exploiting the two U(1) symmetries of the Hamiltonian \eqref{eq:iime_coupling}, i.e. conservation of the total particle number and the $z$-component of the total spin. We compute the effective edge-spin coupling $J^\prime$, given at strong $J$ by the spin gap between singlet and triplet states
\begin{align}
 \Delta_s = E_0 (N, S_\text{tot}=1) - E_0 (N, S_\text{tot}=0)
 .
\end{align}
$E_0 (N,S_\text{tot})$ is the ground state in the Hamiltonian block with total particle number $N$ and total spin $S_\text{tot}$, which can be addressed by selecting either the $S^z_\text{tot}=0$ or the $S^z_\text{tot}=1$ block.

$\Delta_s$ has been calculated for different $J$ between 2.5 and 15 to study the crossover to the strong-coupling regime, where $H_\text{eff}^\prime$ is suggested to be valid and where $\Delta_s=J^\prime$. The results are shown in Fig. \ref{fig:gap} as functions of the edge-spin distance $d^\prime$ (dots connected by solid lines). Indeed, one finds that the asymptotic behavior of the dependence of $\Delta_s$ on the edge-spin distance $d^\prime$ is given by a power-law
\begin{align}
   \Delta_s \sim (d^\prime)^{-\eta}
   .
\end{align}
For $J=15$ and $J=10$ and $d^\prime>7$, $\eta$ has almost converged to $\eta=1$ (see black dashed line) from below. This is a strong evidence for the validity of the suggested two-spin IIME model (Eq. \eqref{eq:eff_ham_iime_offr}) and indicates that the effective on-site interactions at $i=3$ and $i=L-2$ have negligible influence. Decreasing $J$ for fixed $d^\prime$ drives the system out of the IIME limit, and consequently also $\eta$ departs from the predicted value of $1$, e.g. $\eta\rightarrow1.9$ for $J=5$ and $d^\prime>19$. 

Additional insights can be obtained by comparing the numerical results to the spin gaps from perturbation theory (Eq. \eqref{eq:delta_s_offr_pt}), which are shown as colored dashed lines in Fig. \ref{fig:gap}. 
The distance dependence is well recovered by the DMRG results for stronger couplings $J\geq10$. While this is also the case at weaker couplings such as $J=7.5$ for small $d^\prime$, one observes increasing deviations from the perturbative results with increasing system size. 
This can be attributed to Friedel oscillations induced by the open boundaries of the system which lead to a strongly site-dependent local density of states near the system boundaries. 
In particular, the local density of states near the Fermi energy at the impurity positions $i_1=2$ and $i_2=L-1$ strongly decreases with increasing system size $L$. 
Since the Kondo temperature $T_K$ is a very sensitive function of the local density of states, $T_K$ likewise decreases. This tends to increase the Kondo screening cloud and makes the starting point of our analysis based on local Kondo clouds progressively worse.

Our picture is also supported by the $J$-dependence of $\Delta_s$, displayed in the inset of Fig. \ref{fig:gap}. In line with the expectations from the effective two-spin model (Eq. \eqref{eq:eff_ham_iime_offr}), we find that the spin gap is described by $\Delta_s\sim \alpha^2\sim J^{-6}$ for large $J$ with increasing deviations from this relation for smaller $J<10$.

At weak coupling strengths $J\rightarrow0$ the model (Eq. \eqref{eq:ham}) is in the RKKY regime, which also gives rise to an effective two-spin model, yet composed of the two impurity spins $\mathbf S_m$. Their antiferromagnetic indirect coupling $J_\text{RKKY}\sim J^2/d$ leads to a singlet ground state and determines $\Delta_s$. On the other hand, the IIME limit (Fig. \ref{fig:model}b and the left panel of Fig. \ref{fig:model}c), characterized by the effective two-edge-spin model \eqref{eq:eff_ham_iime_offr} with $\Delta_s= J^\prime\sim \alpha^2/d^\prime$, is realized for strong couplings $J\geq10$ as discussed above. Only then Kondo clouds are sufficiently local to induce stable local moments at the chain ends. The crossover region between these regimes, however, lacks a formulation as an effective two-spin model. It can rather be described as a Fermi liquid with vanishing spin gap, with two separate Kondo screening clouds, and a paramagnetic region in-between.

\section{Ferromagnetic distances}
\label{sec:ferro_dist}

\begin{figure}[t]
\centerline{\includegraphics[width=0.95\columnwidth]{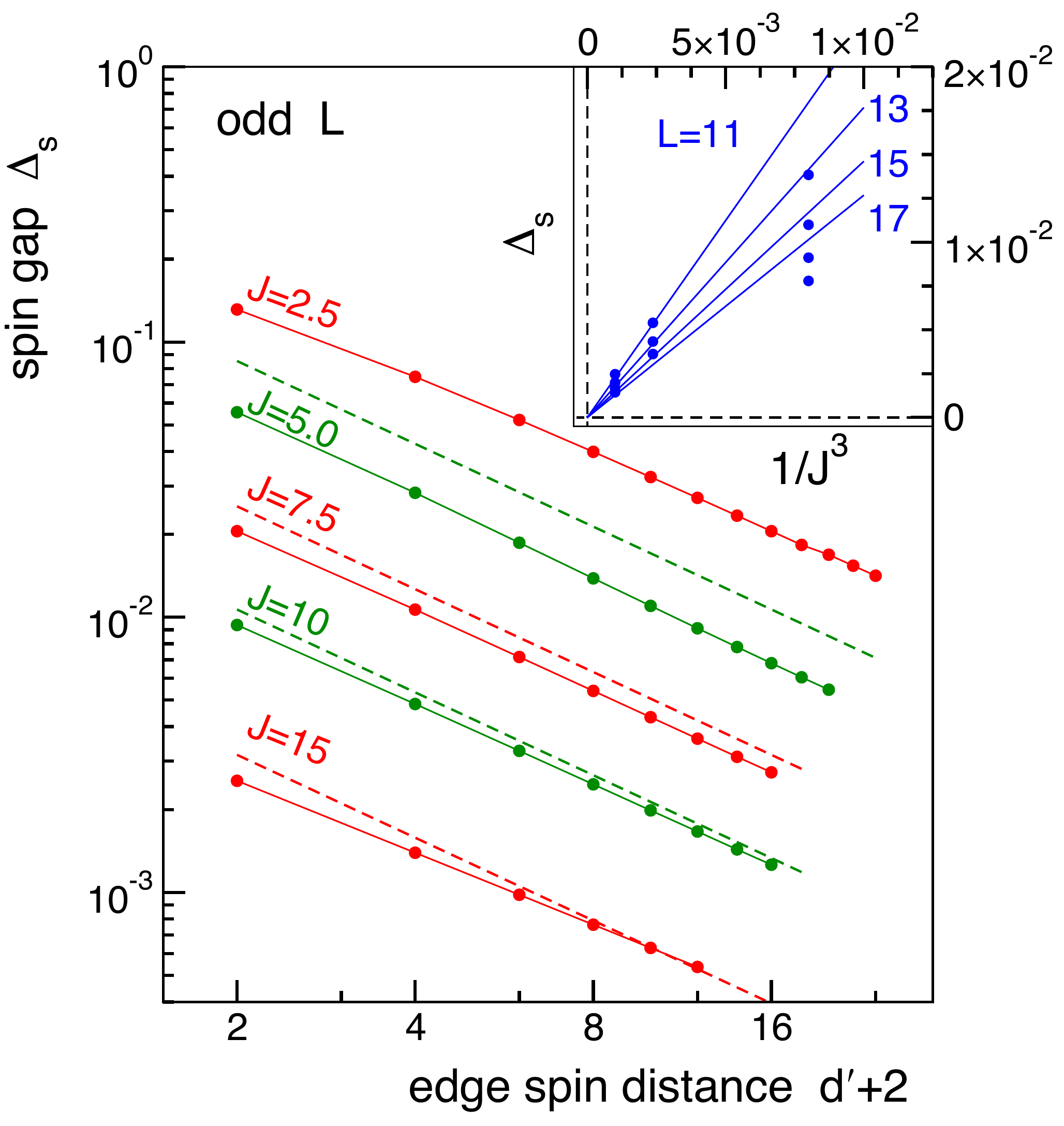}  }
\caption{(Color online)
Spin gap $\Delta_{\rm s}$ as a function of the edge spin distance $d^\prime$ on a double-logarithmic scale for different couplings $J$ as indicated. 
Dots: DMRG results for odd $L$ and even edge-spin distance $d^\prime=L-5$, i.e.\ ferromagnetic coupling. Colored dashed lines: Spin gap as given by first-order perturbation theory in $\alpha/t$, see Eq. \eqref{eq:delta_s_onr_pt}. Black dashed line: Expected $d^\prime$-dependence of $\Delta_s$ for large $d^\prime$ and large $J$.
Inset: $J$-dependence of the spin gap. The DMRG data are approximately described by the expected power law $\Delta_{\rm s} \propto 1/J^{3} \propto \alpha$ at strong $J$.
The nearest-neighbor hopping $t=1$ fixes the energy scale.
}
\label{fig:gap_fm}
\end{figure}

We complete our study by investigating the emerging low-energy model for odd lattice sizes $L$, i.e. even $d^\prime$ (see right panel of Fig. \ref{fig:model}c). Clearly, the low-energy spectrum of the non-interacting central region, which builds up at strong $J$, now contains a singly occupied Fermi level due to the odd number of sites. Performing the same type of perturbation theory as above for weak coupling $\alpha\ll t$, shows that the effective ferromagnetic RKKY interaction between the two edge spins is exceeded by a linear-in-$\alpha$ coupling between single edge spins and the spin of the delocalized Fermi electron \cite{SGP12} of the form
\begin{align}
    \label{eq:eff_ham_iime_onr}
    H_\text{eff}^\prime = -J^\prime (\mathbf s_1 + \mathbf s_L) \mathbf s_F
    .
\end{align}
The coupling is ferromagnetic and given by
\begin{align}
    \label{eq:delta_s_onr_pt}
    J^\prime = \alpha |U_{3,k_F}|^2
             = \alpha |U_{L-2,k_F}|^2 
             = 2\alpha /(d^\prime+2)
    .
\end{align}
Thus, rather than an RKKY model, the system renormalizes to a central-spin model \cite{G76}, as sketched in the right panel of Fig. \ref{fig:model}c, with the delocalized spin of the Fermi electron $\mathbf s_F$ as the central spin. The ground state of the ferromagnetic central-spin model is a total spin quartet ($S_\text{tot}=3/2$).

The $J$ as well as the distance dependence of the effective coupling constant $J'$ can be checked using the DMRG by computing the gap
\begin{align}
 \Delta_s = E_0(N,S_\text{tot}=1/2) - E_0(N,S_\text{tot}=3/2)
 , 
\end{align}
see Fig. \ref{fig:model}d (right panel).
Note that the spin gap of the central-spin model is related to $J'$ as $\Delta_s=J^\prime/2$.
To get $\Delta_s$ as a difference of two ground-state energies in sectors with different $S_{\text{tot}}$ but possibly equal $S^{z}_{\text{tot}}$, the DMRG algorithm must be extended slightly.
Making use of the matrix-operator representation of the Hamiltonian, we have implemented an additional interaction term
\begin{align}
  H \mapsto H + \lambda(\mathbf S^2_\text{tot} - S_\text{tot}^\text{target}(S_\text{tot}^\text{target}+1))^2 \: .
\end{align}
For $\lambda>0$, this allows us to target blocks with a given total-spin quantum number $S_\text{tot}^\text{target}$.

Our calculations show that there is an incidental degeneracy of the spin-quartet ground state of the central-spin model with a spin-doublet state which is different from the excited $S_\text{tot}=1/2$ state of the central-spin model.
This doublet, however, can be shifted to higher energies by adding a weak repulsive Hubbard term, $U(n_\uparrow-\frac12)(n_\downarrow-\frac12)$ to the conduction-electron system. 
Namely, the Lieb-Mattis theorem \cite{She96,Tsu97a} dictates that the total ground-state spin of the {\em correlated} ($U>0$) model must have $S_\text{tot}=3/2$.
We have used weak Hubbard interaction strengths of $U=0.05$ up to $0.25$ and have checked that this does not significantly affect the results obtained for $\Delta_{s}$.

The results in Fig. \ref{fig:gap_fm} indicate that the linear $\alpha$ dependence of $\Delta_s$, suggested by perturbation theory (Eq. \eqref{eq:delta_s_onr_pt}), is indeed realized at strong couplings $J\geq10$ (see also inset of Fig. \ref{fig:gap_fm}). This behavior along with the considerably larger gaps distinguishes the ferromagnetic case clearly from the antiferromagnetic case discussed in the preceeding sections.

There is also a remarkable agreement between the numerical and the perturbative data with regard to the distance dependence of $\Delta_s$ (Fig. \ref{fig:gap_fm}). 
Unlike the antiferromagnetic case, even at relatively small couplings such as $J=5$ and for all considered system sizes $L$, $\Delta_s$ scales as $1/(d^\prime + 2)$, see Eq.\ (\ref{eq:delta_s_onr_pt}).  
Larger energy gaps make the effective central-spin model in the ferromagnetic case more robust as compared to the effective low-energy model in the antiferromagnetic case.

For extremely large system sizes $L$ (not accessible here) and for still strong enough $J$, we expect that the finite-size physics, emerging here as effective central-spin model, is replaced by a conventional RKKY interaction obtained within the thermodynamic limit \cite{SGP12}. 
The reason is that the linear-in-$\alpha$ contribution to the effective low-energy Hamiltonian will be less and less important with increasing $L$ as compared to higher-order terms, which indicates the breakdown of finite-size perturbation theory.

\section{Conclusions}

Quantum-confined multi-impurity Kondo systems exhibit a complex interplay of different magnetic exchange mechanisms competing or cooperating with the Kondo effect. 
Our present study has addressed a prototypical model where the emergence of local bound states in the strong Kondo-coupling regime ($J\gg t$), i.e.\ the formation of local Kondo singlets, strongly confines the conduction-electron mobility.
We have demonstrated that this leads to the formation of local spin moments in the conduction-electron system. 
For the one-dimensional system studied here, moments are created at the edges of the chain. 
Virtual excitations of the local Kondo singlets mediate an indirect coupling $\alpha \propto 1/J^{3}$ of these edge spins to the local magnetic moments at the edges of the remaining one-dimensional conduction-electron system. 
This inverse indirect magnetic exchange (IIME) is the strong-coupling analog of the well-known indirect RKKY exchange which operates at weak $J\ll t$.
An interesting observation made here is that the central part of the conduction-electron system can mediate an effective mutual coupling of the edge spins which can be understood by {\em weak-coupling} RKKY-like perturbation theory as for $J\to \infty$ the effective IIME coupling $\alpha \to 0$ is weak.

As evidenced by numerical DMRG results and by analytical insights from strong-coupling (in $J$) and weak-coupling (in $\alpha$) perturbation theory, we have seen that odd-even effects are crucial to understand the total ground-state spin, the elementary excitation gap and the distance as well as the $J$ dependence of the effective magnetic coupling between the edge spins.
Namely, depending on the number of lattice sites $L$, two very different effective low-energy models are obtained:
For odd distances $d^\prime = L-5$ between the edge spins, an effective antiferromagnetic two-spin model emerges which is composed of the two edge spins and coupled by an effective RKKY interaction $\propto \alpha^{2}$. 
In this case, where RKKY emerges from IIME, the spin gap is given by $\Delta_s \sim \alpha^2 / d^\prime$.
For even distances, on the other hand, we find an effective central-spin model where the two edge spins couple ferromagnetically to the delocalized spin of the electron in the highest occupied spin-degenerate one-particle state of the conduction-electron system.
This induces ferromagnetic correlations between the two edge spins, and the ground state is essentially given by a spin quartet, but there is no indirect ferromagnetic coupling, i.e.\ this case is fundamentally different from a ferromagnetic RKKY model. 
This is also reflected in a spin gap $\Delta_s = \alpha/(2(d^\prime+2))$ which is linear rather than quadratic in $\alpha$.

Quantum confined Kondo systems with a wide range of accessible model parameters can be simulated by ultracold Fermi atoms trapped in optical lattices. 
We believe that the systematic study of magnetic exchange interactions and Kondo correlations in those systems is only at the very beginning. 
The largely different physics of effective low-energy models that are obtained by slightly different geometries or slightly different impurity-spin configurations offers an exciting perspective for tailoring physical properties in the experiment.
Further theoretical work may e.g.\ address the low-energy physics of stacked Kondo singlets where the number of intermediate Kondo singlets between two edge spins is varied. 
Quite generally, not much is known on the crossover from a single or few Kondo impurities to the dense Kondo lattice in one- and higher-dimensional quantum-confined systems.

\acknowledgments

Financial support of this work by the Deutsche Forschungsgemeinschaft within the SFB 925 (project B5) 
is gratefully acknowledged.

\end{document}